\documentstyle[aps,epsfig,prd]{revtex}
\newcommand{\cd}{\cdot}

\newcommand{\al}{\alpha}

\newcommand{\de}{\delta}
\newcommand{\De}{\Delta}

\newcommand{\ga}{\gamma}
\newcommand{\Ga}{\Gamma}

\newcommand{\la}{\lambda}
\newcommand{\Om}{\Omega}
\newcommand{\om}{\omega}
\newcommand{\si}{\sigma}

\newcommand{\ra}{\rightarrow}

\newcommand{\be}{\begin{equation}}
\newcommand{\ee}{\end{equation}}
\newcommand{\gsim}{\stackrel{>}{\sim}}
\newcommand{\lsim}{\stackrel{<}{\sim}}
\newcommand{\bea}{\begin{eqnarray}}
\newcommand{\eea}{\end{eqnarray}}
\newcommand{\bean}{\begin{eqnarray*}}
\newcommand{\eean}{\end{eqnarray*}}

\newcommand{\bk}{{\bf k}}
\newcommand{\bq}{{\bf q}}
\newcommand{\bp}{{\bf p}}
\newcommand{\bx}{{\bf x}}

\begin{document}
\draft
\preprint{\
\begin{tabular}{rr}
&
\end{tabular}
}
\twocolumn[\hsize\textwidth\columnwidth\hsize\csname@twocolumnfalse\endcsname
\title{Gravitational wave production: A strong constraint on
  primordial magnetic fields}
\author{ Chiara Caprini$^{1,2}$ and Ruth Durrer$^1$}
\address{ $^1$D\'epartement de Physique Th\'eorique, Universit\'e de
Gen\`eve,
24 quai Ernest Ansermet, CH-1211 Gen\`eve 4, Switzerland\\
$^2$Dipartimento di fisica, Universit\`a degli Studi di Parma, Parco Area
delle Scienze 7A, 43100 Parma, Italy}
\maketitle

\begin{abstract}
We compute the gravity waves induced by anisotropic stresses of
stochastic primordial magnetic fields. The nucleosynthesis bound on
gravity waves is then used to derive a limit on the magnetic field
amplitude as function of the spectral index. The obtained limits are
extraordinarily strong: If the  primordial magnetic field is produced by
a causal process, leading to a spectral index $n\ge 2$ on super
horizon scales, galactic
magnetic fields produced at the electroweak phase transition or
earlier have to be weaker than $B_\la \le 10^{-27}$Gauss! If they are
induced during an inflationary phase (reheating temperature
$T\sim 10^{15}$GeV)  with a spectral index $n\sim 0$, the magnetic
field has to be weaker than  $B_\la \le 10^{-39}$Gauss! Only very red
magnetic field spectra, $n\sim -3$ are not strongly constrained.
We also find that a considerable amount of the magnetic field energy
is converted into  gravity waves.

The gravity wave limit derived in this work rules out most of the
proposed processes for primordial  seeds for the
large scale magnetic fields observed in galaxies and clusters.
\end{abstract}
\date{\today}
]
\renewcommand{\thefootnote}{\arabic{footnote}} \setcounter{footnote}{0}

\section{Introduction}
Our galaxy, like most other spiral galaxies, is permeated by a magnetic
field of the order of $B\sim 10^{-6}$Gauss.
Recently, similar magnetic fields have also been
observed in clusters of galaxies on scales of up to
$\la\sim 0.1$Mpc~\cite{kronberg,Eil99}. There is an ongoing debate
whether such
fields can be produced by charge separation processes during
galaxy and cluster formation~\cite{astro} or whether primordial seed
fields are needed, which have then been amplified by simple
adiabatic contraction or by a dynamo mechanism. In the first case,
seed fields of $B\sim 10^{-9}$Gauss are needed while in the second case
$B\sim 10^{-20}$Gauss~\cite{astro} or even 10$^{-30}$Gauss
in a universe with low mass density~\cite{davis} suffice.
Several mechanisms have been proposed for the origin
of such seed fields, ranging from inflationary production of magnetic
fields~\cite{inflation,Ann,stringcos} to cosmological phase 
transitions~\cite{phase}.

Primordial magnetic fields have been constrained in the past in
various ways mainly by using their effect on anisotropies in the
cosmic microwave background~\cite{jenni,kosowsky,evan,RTA,BFS,RPT}. In
these works constant magnetic fields and stochastic fields with red
spectra $n\sim -3$~\cite{RPT} have been considered and the limits
obtained where of the order of a
few$\times 10^{-9}$Gauss. A simple order of magnitude estimate shows
that, from the CMB alone, one cannot expect much stronger constraints of 
magnetic fields: The energy density in a magnetic field is
\be
\Om_B = {B^2\over 8\pi\rho_c} \simeq 10^{-5}\Om_{\ga}
   (B/10^{-8}{\rm Gauss})^2~,
\ee
where $\Om_{\ga}$ is the density parameter in photons.
We naively expect a magnetic field of $10^{-8}$Gauss  to
induce perturbations in the CMB on the order of $10^{-5}$, which are
just on the level of the observed CMB anisotropies. It is thus
expected that CMB anisotropies cannot constrain primordial magnetic
fields to better than a few tenths of this amplitude.

In this work we constrain magnetic fields by the gravity waves which they
induce classically, via the anisotropic stresses in their energy momentum 
tensor. These gravity waves lead to much stronger 
constraints than CMB anisotropies, especially for spectral
indices $n>-3$. This comes from the fact that the spectrum of the gravity wave 
energy density induced by stochastic magnetic fields is always blue (except for
$n=-3$ where it is scale invariant) and thus leads to stronger
constraints on small scales than on the large scales probed by CMB
anisotropies.

The effects of a constant magnetic field on gravity wave
evolution and production have been studied in~\cite{Roy}. Here we
concentrate on the production of gravity waves, but consider a
stochastic magnetic field.

The remainder of this paper is organized as follows: In the next
section we define the initial magnetic field spectrum and its
evolution in time, and we  determine the magnetic stress tensor which
sources gravity waves. In Section~3 we calculate the induced gravity
wave spectrum
and estimate the effect of back-reaction. In Section~4 we derive
limits on the primordial magnetic field using the nucleosynthesis
limit on gravity waves and discuss our conclusions. In order not to
loose the flow of the arguments, several technical derivations are 
deferred to three appendices.

We use conformal time which we denote by $\eta$; the scale factor is
$a(\eta)$. Derivatives w.r.t conformal time are denoted by an over-dot,
${da \over d\eta} =\dot a$. We normalize the scale factor today to
$a(\eta_0)=1$. The index $0$ on a time dependent variable always indicates
today. We assume a spatially flat universe with vanishing cosmological 
constant throughout. Neglecting a possible cosmological constant 
modifies the evolution of the scale factor only at very late times, 
$z<2$ and is therefore irrelevant for the results of this paper.
We set the speed of light $c=1$ so that times and length scales can be given
in units of sec, cm or Mpc, whatever is convenient. With our conventions,
the scale factor is given by
\be
        a(\eta)=H_0\eta({H_0\eta\over 4}+\sqrt{\Om_{\rm rad}})~, 
\label{scfac}
\ee
where $ H_0 =(3.086\times 10^{17}{\rm sec})^{-1}h_0$ is the Hubble 
parameter,
$0.5<h_0<0.8$ and $\Om_{\rm rad}=4.2\times 10^{-5}h_0^{-2}$ is the radiation
density parameter (photons and three types of massless neutrinos).

Note that the scale factor has no units,
but conformal time and comoving distance do. The normalization of $a$ 
implies
that comoving distance becomes physical distance today. The conformal time
$\eta$ is the comoving size of the horizon. The relation between $\eta$ and
redshift or temperature is simply
\bea 
	z(\eta) &= & {1\over a(\eta)}-1 ~,\nonumber \\
   T(\eta) &=& z(\eta) T_0 \simeq  z(\eta)2.4\times 10^{-4}{\rm eV}~. 
\label{rela}
\eea
The comoving time of equal matter and radiation, defined by
$a(\eta_{eq})^{-3} = \Om_{\rm rad}a(\eta_{eq})^{-4}$ or
$z_{eq}+1= \Om_{\rm rad}^{-1}$, is
\be \eta_{eq} =2(\sqrt{2}-1)                            \label{equal}
        \sqrt{\Om_{\rm rad}}H_0^{-1}\sim 1.7\times10^{15}{\rm sec}.
\ee
Greek indices run from $0$ to $3$, Latin
ones from $1$ to $3$. Spatial (3d) vectors are denoted in bold.

\section{Primordial stochastic magnetic fields}
In this section we closely follow Ref.~\cite{RPT}.
During the evolution of the universe, the
conductivity of the inter-galactic medium is effectively infinite.
We can decouple the time evolution of the magnetic field from its
spatial structure:
{\bf B} scales like $B^2(\eta,{\bf x})={\bf B}_0^2({\bf x})/a^4$ on
sufficiently large scales. (In our coordinate basis $B_i\propto 1/a$
and $B^i\propto a^{-3}$ as can be derived easily from Maxwell's equations in
curved spacetime with vanishing electric field, see e.g.~\cite{DS}).
On smaller scales, the interaction of the
magnetic field with the cosmic plasma becomes important, leading mainly
to two effects: on intermediate scales, the field oscillates like
$\cos(v_Ak\eta)$, where $v_A = (B^2/(4\pi(\rho+p)))^{1/2}$ is the
Alfv\'en velocity, and on very small scales, the field is exponentially
damped due to shear viscosity~\cite{mhd,sblong,chiara}. We will take into
account the time dependent damping scale as a time  dependent cutoff
$k_d(\eta)$ in the spectrum of ${\bf B}$. As we shall see, our
constraints  come  from  small scales
where the spectrum is exponentially damped and
oscillations can be ignored. We therefore disregard them in what follows.
The expressions for $k_d(\eta)$ are derived in Appendix~\ref{Apdamp}.
The only result of this appendix relevant here is that the damping scale
$1/k_d(\eta)$ grows like a positive power $\al>0$ of $\eta$ and is 
always smaller than the horizon scale, $k_d(\eta)\propto 1/\eta^\al$ and
$k_d(\eta)>1/\eta$. The reader not interested in the details of damping and
confident with this relatively obvious result, can skip 
Appendix~\ref{Apdamp}.

We model ${\bf B}_0({\bf x})$ as a statistically homogeneous
and isotropic random field. The transversal nature of ${\bf B}$
then leads to
\be
  \langle B^i({\bf k})B^{*j}({\bf q})\rangle =
        \de^3({\bf k-q})(\de^{ij}-\hat{k}^i\hat{k}^j)B^2(k)~. \label{powB}
\ee
We use the Fourier transform conventions
\bean
\ B^j({\bf k}) &=&\int d^3x
\exp(i{\bf x\cd k})B_0^j({\bf x}) ~,\\
  B_0^j({\bf x}) &= & {1\over (2\pi)^3}\int d^3k \exp(-i{\bf x\cd k})
         B^{j}({\bf k}) ~,
\eean
and $\hat{\bf k} ={\bf k}/k$, $k =\sqrt{\sum_i(k^i)^2}$;
$\bk$ is the wave vector today which is also the co-moving wave vector. Its
unit is inverse length which we will express in sec$^{-1}$.

We want to derive a limit on the amplitude of magnetic fields on the scale
 $\la \sim 0.1$Mpc generated by a primordial process which took place before
$\eta =0.1{\rm Mpc}\sim 10^{13}{\rm sec}$ corresponding to $T\sim 1$keV.
Hence we are mainly interested in magnetic fields generated on super horizon
scales. As we shall see, our limits only apply for fields generated before
nucleosynthesis, $T>T_{nuc}\simeq 0.1$MeV. The main examples we have in 
mind are inflationary generation of magnetic fields~\cite{inflation,Ann}, 
magnetic fields generated
in string cosmology~\cite{stringcos} and magnetic fields generated during
the electroweak phase transition~\cite{phase}.

In the first two examples, a simple power law magnetic field spectrum with
upper cutoff $k_c\simeq \eta_{in}^{-1}$ is generated. The conformal time
$\eta_{in}$ marks the end of inflation or the string scale respectively.

Electroweak magnetic field production is causal, leading mainly to fields on
scales smaller than the size of the horizon at the phase transition,
$\eta_{ew}\simeq 4\times10^4 $sec $\simeq 10^{15}$cm
  $\simeq 3\times 10^{-4}$pc.
These sub-horizon fields, which cannot propagate into larger scales during
the linear evolution discussed in this paper, and which are essentially damped 
by viscosity, will be neglected in this paper. Motivated from inflation, we 
simply impose an initial cutoff scale $k_c(\eta_{in})=1/\eta_{in}$.
Allowing for more small scale power, as it is certainly present initially in
causal mechanisms, only strengthens our result which actually comes from the
smallest scales not affected by damping.

If $\bf B$ is generated by a {\it causal} mechanism, it is
uncorrelated on super horizon scales,
\be
\langle B^i({\bf x},\eta)B^j({\bf x}',\eta)\rangle=0  ~~~{\rm for }~~~
|{\bf x}-{\bf x}'| > 2\eta ~.  \label{causal}
\ee
Here it is important, that the universe is in a stage of standard
Friedman  expansion, so that the comoving causal horizon size is about
$\eta$.
During an inflationary phase, the causal horizon diverges and our
subsequent argument does not apply. In this somewhat misleading sense,
one calls inflationary perturbations 'a-causal'.

According to Eq.~(\ref{causal}),
$ \langle B^i({\bf x},\eta)B^j({\bf x}',\eta)\rangle$ is a
function with compact support and hence its Fourier
transform is analytic. The function
\be
  \langle B^i({\bf k})B^{*j}({\bf k})\rangle \equiv
(\de^{ij}-\hat{k}^i\hat{k}^j)B^2(k)
\ee
is analytic in {\bf k}. If we assume also that $B^2(k)$ can be
approximated by a simple power law, we must conclude that $B^2(k)
\propto k^n$, where $n\ge 2$ is a even integer. (A white noise
spectrum, $n=0$ does not work because of the transversality condition
which has led to the non-analytic pre-factor $\de^{ij}-\hat{k}^i\hat{k}^j$.)
By causality, there can be no deviations from this law on scales larger
than the horizon size at formation, $\eta_{in}$. As explained above, we
neglect fields on smaller scales by a simple cutoff.

We assume that ${\bf B}_0$ is a Gaussian random field. Although this is 
not the most general case, it greatly simplifies calculations 
and  gives us a good idea of what to expect in more general situations.

Using Wick's theorem for Gaussian fields we can calculate
the correlator of the tensor contribution to the anisotropic stresses 
induced  by the magnetic field, which  we denote by $\Pi_{ij}$.  One finds 
(see Appendix~\ref{Apten})
\bea
\langle \Pi^{ij}({\bf k},\eta)\Pi^{*lm}({\bf
        k}',\eta) \rangle &=&|\Pi(k,\eta)|^2/a^{12}{\cal M}^{ijlm}\de({\bf
k-k'})\nonumber \\
\langle \Pi_{ij}({\bf k},\eta) \Pi^{*ij}({\bf k'},\eta)\rangle
&=& {4\over a^8}f^2(k,\eta)\de({\bf k-k'}),
\label{powPi}
\eea
where
\bea
{\cal M}^{ijlm}(\bk)&\equiv&\de^{il}\de^{jm}+\de^{im}\de^{jl}   
	-\de^{ij}\de^{lm}+ k^{-2}(\de^{ij}k^lk^m + \nonumber \\
&&      \de^{lm}k^ik^j -\de^{il}k^jk^m - \de^{im}k^lk^j -\de^{jl}k^ik^m
        \nonumber \\
&&      -\de^{jm}k^lk^i) + k^{-4}k^ik^jk^lk^m~,
\eea
and
\be
f(k)^2 ={1\over 16(2\pi)^8} \int d^3 q B^2(q) B^2(|{\bf k} - {\bf q}|)
(1+2\gamma^2+\gamma^2 \beta^2) ~, \label{ff}
\ee
with $\gamma={\bf \hat k}\cdot{\bf \hat q}$
and  $\beta = {\bf \hat k}\cdot { \widehat{\bf k-q}}$.
For this result we made use of statistical isotropy, which implies that the two
spin degrees of freedom of $\Pi_{ij}$ have the same average amplitude.
More explicitly: in a coordinate system where $\bk$ is parallel to the
$z$-axis,  $\Pi_{ij}$ has the form
\[
\left(\Pi_{ij}\right) = \left(\begin{array}{ccc}
   \Pi_+ & \Pi_\times & 0 \\
  \Pi_\times & -\Pi_+ & 0 \\
   0 & 0 & 0 \end{array} \right)~;
\]
together with Eq.~(\ref{powPi}), statistical isotropy then gives
\be \langle|\Pi_+|^2\rangle =
\langle|\Pi_\times|^2\rangle =  {1\over a^4}f^2
  \label{modes}~.
\ee
To continue, we have to specify $B^2(k)$. For simplicity we assume a
simple power law with cutoff $k_c$ which can depend on time. As all scales
smaller than $1/ k_d(\eta)$ are damped, clearly we have to require
$k_c(\eta)\le k_d(\eta)$. Motivated by inflationary magnetic
field production we choose $k_c(\eta_{in}) \sim 1/\eta_{in}$, the
primordial magnetic field is coherent up to the horizon size at
formation. For magnetic fields produced during the electroweak phase
transition, the 'coherence scale' is substantially smaller~\cite{Ola},
$k_c(\eta_{in}) \gg 1/\eta_{in}$  which would
strengthen our limit as we shall see. Since it is unphysical to assume
$k_c(\eta_{in})< 1/\eta_{in}$, our assumption is conservative.
 We set
\[ k_c(\eta)=\min( 1/\eta_{in}, k_d(\eta)) ~.\]
It is important to keep in mind, that this cutoff scale
is always smaller than the horizon scale.

We now can parameterize $B^2$ by
\be
B^2(k)= \left\{\begin{array}{ll}
        {(2\pi)^5\over 2}
{(\lambda/\sqrt{2})^{n+3}\over \Gamma[\frac{n+3}{2}]}B_{\lambda}^2k^n &
        \mbox{ for } k<k_c\\
        0 &     \mbox{ otherwise.} \end{array} \right.
\label{magfield}\ee
The normalization is such that
\be
B_\la^2 ={1\over V}\int d^3r \langle B_0(\bx)B_0(\bx +{\bf r})\rangle
    \exp(-{r^2\over 2\la^2})~,
\ee
where $V=\int d^3r\exp(-r^2/2\la^2) =\la^3(2\pi)^{3/2}$ is the
normalization volume.
(We have assumed that the cutoff scale is smaller than
$\lambda$.) We will finally fix $\la =0.1h^{-1}$Mpc, the largest scale on
which coherent magnetic fields have been observed; but the scaling of
our results with $\la$ will remain obvious.

The energy density in the magnetic field at some arbitrary scale
$\ell$ is $\propto B_{\ell}^2 \propto B^2(k)k^3|_{k=1/\ell} \propto
\ell^{-(n+3)}$.
In order not to over-produce long range coherent fields, we  must
require $n\ge -3$. For $n=-3$ we obtain a
scale invariant magnetic field energy spectrum.

Using Eqs.~(\ref{magfield}) and (\ref{ff}) we can calculate $f$. The
integral cannot be computed analytically, but the following result is
a good approximation for all wave numbers $k$~\cite{RPT}
\bea
  f^2(k,\eta) &\simeq& A  \times  \left\{\begin{array}{ll}
        k_c(\eta)^{2n+3}  & \mbox{ for } n \ge -3/2 \\
        k^{2n+3}   & \mbox{ for } n \le -3/2 ~. \end{array} \right.
\label{faprox}\\
\mbox{with} && \nonumber \\
A &=& {(2\pi)^3\over 16}
                \frac{(\lambda/\sqrt{2})^{2n+6}B_\lambda^4}
{\Gamma^2[\frac{n+3}{2}]}  \nonumber
\eea
For  $n>-3/2$, the gravity wave source $\Pi$ is  white noise,
independent of $k$. Only the amplitude, which is proportional to
$(\la k_c)^{2n}$, depends on the spectral index. This is due to the
fact that the integral (\ref{ff}) is dominated by the contribution from
the  smallest scale $k_c^{-1}$. The induced  gravity wave  spectrum
will therefore be a white noise spectrum for all $n>-3/2$.

\section{Gravity waves from magnetic fields}
We now proceed to calculate the gravity waves induced by the magnetic
field stress tensor.
The metric element of the perturbed Friedman universe is given  by
\[ds^{2}=a^{2}(\eta)[d\eta^2-(\delta_{ij}+2h_{ij})
dx^idx^j]~, \]
where $h^i_i=0$ and $h^j_{i}k^{i}=0$ for tensor perturbations.
The magnetic field sources the evolution  of $h_{ij}$
through
\be
{\ddot{h}}_{ij}+2{\dot a \over a}\dot{h}_{ij}+
k^2 h_{ij}=8\pi G \Pi_{ij} ~.
\label{hprime}
\ee
$ \Pi_{ij}$ is a random variable, but its time evolution is deterministic, it
evolves in time simply by redshifting and by the
evolution of the cutoff. Each component is given by
\[
\Pi_{\bullet} (k,\eta) = {1\over a^2}f(k,\eta)
        \tilde\Pi_{\bullet}(k)~,
\]
where $\tilde\Pi_{\bullet}(k)$ is a time independent random variable with
power spectrum $\langle|\tilde\Pi_{\bullet}(k)|^2\rangle =1$. Therefore,
also each component of the induced gravity wave is given by
\[ h_{\bullet} (k,\eta) = h(k,\eta)\tilde{\Pi}_{\bullet}(k)~, \]
where $h(k,\eta)$ is a solution of
\be
\ddot h+2{\dot a \over a}\dot h +
k^2h={8\pi G\over a^2(\eta)}f(k,\eta) ~.
\label{Hprime}
\ee
The gravity wave power spectrum is then given by
\be
\langle \dot{h}^{ij}(\bk,\eta) \dot{h}^{*}_{ij}(\bk',\eta)\rangle
   = 4\dot{h}^2(k,\eta)\de(\bk-\bk')~.  \label{hk}
\ee

In real space ,the energy density in gravity waves is
\[
   \rho_G = {\langle \dot{h}_{ij} \dot{h}^{ij}\rangle \over 16\pi Ga^{2}}~.
\]
The factor $1/a^2$ comes from the fact that $\dot
h$ denotes the derivative w.r.t. conformal time. Fourier transforming
this relation, we obtain with Eq.~(\ref{hk})
\be
  \rho_G
  = \int_0^{k_c}{dk\over k} {d  \rho_G(k) \over d\log (k)}~, \label{rhoG}
\ee
with
\[
{d  \rho_G(k) \over d\log (k)} =  {k^3\dot h^2 \over a^2(2\pi)^6G}
\]
such that
\be
{d\Om_G(k)\over d\log(k)} \equiv  {d  \rho_G(k) \over \rho_c\,d\log (k)} =
{k^3\dot h^2 \over a^2\rho_c(2\pi)^6G}~, \label{Omgk}
\ee
where  $\rho_c=3H_0^2/(8\pi G)$ denotes the critical density today.
In Appendix~\ref{apwave} we solve Eq.~(\ref{Hprime}) for $n<-3/2$, when $f$ 
is time independent, and we show that for wave
numbers which enter the horizon in the radiation dominated era, the density
parameter in gravity waves produced by the magnetic field can be expressed 
as
\bea
&& {d\Om_G(k)\over d\log(k)} \simeq
%{4k^3f(k)^2(8\pi G)^2\log^2(x_{in})\over H_0^4\Om_{\rm rad}3(2\pi)^5} =
{12k^3f(k)^2\log^2(x_{in}) \over \rho_c^2\Om_{\rm rad}(2\pi)^5}~, 
  \label{OmGk1}\\
&& \mbox{ for }~ n\le -3/2 ~. \nonumber
\eea
Fourier transforming the expression for the  magnetic field energy
$\rho_B = \langle B^2(\bx)\rangle/(8\pi)$, we obtain the magnetic field 
density parameter at time $\eta$,
 \bea
{d\Om_B(k)\over d\log(k)} &=& {B_\la^2\over 8\pi\rho_c}{(k\la)^{n+3}\over
2^{(n+3)/2}\Ga({n+3\over 2})} \label{OmBk}\\
\Om_B(\eta)= \Om_B(k_c(\eta)) &=& \int_0^{k_c(\eta)}{dk\over k}{d\Om_B(k)\over d\log(k)}
\nonumber\\
  &=& {B_\la^2\over 8\pi\rho_c}
{(k_c\la)^{n+3}\over 2^{(n+5)/2}\Ga({n+5\over 2})}~.
\label{OmB}
\eea
Note that $\Om_B$ may well be considerable on small scales, since this is 
the magnetic field energy at very early times which can be damped and
transformed, e.g. into radiation later. But of course, for our perturbative 
calculation to apply, we must require ${d\Om_B(k)\over d\log(k)} <
 \Om_{\rm rad}$ during the radiation dominated era.
%Therefore, we do not have a
%significant observational limit on ${d\Om_B(k)\over d\log(k)}$ for very 
%large
%wave numbers. Clearly we require  ${d\Om_B(k)\over d\log(k)}< 1$.
Using Eqs.~(\ref{OmBk},\ref{OmB}) and the result~(\ref{faprox})
for $f$, we obtain from Eq.~(\ref{OmGk1})
\bea
{d\Om_G(k)\over d\log(k)}
  &\simeq& {({d\Om_B(k)\over d\log (k)})^2\over
        \Om_{\rm rad}}24\log^2(x_{in})~,
\label{nl32k} \\
&& \mbox{ for } -3< n<-3/2  \nonumber  \\
\Om_G  &=& \int_0^{1/\eta_{in}}{dk\over k}{d\Om_G(k)\over
d\log(k)} \nonumber \\
&\simeq& {\Om_B^2(\eta_{in})\over \Om_{\rm rad}}12(n+3)
                ~, \label{nl32I} \\
&&  \mbox{ for } -3< n<-3/2~. \nonumber
\eea
In the integrated formula for $\Om_G$ we have neglected the  logarithmic
dependence $\log^2(x_{in})$.

If $n>-3/2$ the result changes since $f$ now depends on
time via the cutoff $k_c(\eta)=\min(1/\eta_{in},k_d(\eta))$. Clearly,
$k_d(\eta_{in})>1/\eta_{in}$ by causality. We define the time
$\eta_{visc}$ to be the moment when the damping scale becomes smaller
than $\eta_{in}$, $k_d(\eta_{visc}) =1/\eta_{in}$. From that time on,
the function $f$ decays like a power law,
\[  f^2(k,\eta) \propto k_d^{2n+3} \propto
   f^2(k,\eta_{in})(\eta_{visc}/\eta)^{\al(2n+3)}~, \]
where $\al$ is a positive power describing the growth of the viscosity
damping scale. Hence, the source term of Eq.~(\ref{Hprime}) starts
to decay faster than $1/a^2$, and additional gravity wave
production after $\eta_{visc}$ is sub-dominant. We neglect it in our
attempt to derive an upper limit for primordial magnetic
fields. For $n>-3/2$, the gravity wave solution given in
Appendix~\ref{apwave}, Eq.~(\ref{aphx}) is then simply modified by
$-\log(x_{in}) \ra \log(x_{visc}/x_{in})$, since the integral of the
gravity wave source term only has to extend from $x_{in}$ to $x_{visc}$.
Taking  also into account
that up to $\eta_{visc}$ the cutoff scale is $k_c(\eta)=1/\eta_{in}$,
hence $f^2(k,\eta) \propto k_c^{2n+3}=1/\eta_{in}^{2n+3}$, we obtain
\bea
{d\Om_G(k)\over d\log (k)}
  &\simeq& {\left({d\Om_B(k)\over d\log (k)}\right)^2
        (k\eta_{in})^{-3-2n} \over \Om_{\rm rad}}24
        \log^2(x_{visc}/x_{in}),\nonumber \\
&&    \mbox{ for } n>-3/2  \label{ng32k}\\
\Om_G   &\simeq & {\Om_B^2(\eta_{in})\over
  \Om_{\rm rad} }8(n+3)^2\log^2(\eta_{visc}/\eta_{in}) ~ ,  \label{ng32I}\\
&& \mbox{ for } n>-3/2~. \nonumber
\eea
In Appendix~\ref{Apdamp}, we estimate for the two examples of
inflation, $T_{in} \sim 10^{15}$GeV, $\eta_{in}\sim 8\times 10^{-9}$sec and
the electroweak phase transition, $T_{ew} \sim 200$GeV, $\eta_{in} =
 \eta_{ew} \sim 4\times 10^{4}$sec, 
\bean
  \eta_{visc}/\eta_{in} \simeq 10^9 & \mbox{ for inflation } & \eta_{in}=
	8\times 10^{-9}{\rm sec}, \\
  \eta_{visc}/\eta_{ew} \gsim 3000 & \mbox{ for ew. trans. } &
\eta_{ew}\sim 4\times 10^{4}{\rm sec}~.
\eean
Up to logarithms,
the final formula for gravity wave production  is nearly the same
for all values of the spectral index (cf. Eqs.~(\ref{ng32I}) and
(\ref{nl32I})).

In these formulas back-reaction, namely the decrease of magnetic field
energy due to the emission of gravity waves, is not included.
Therefore Eqs.~(\ref{nl32k},\ref{nl32I}) and
(\ref{ng32k},\ref{ng32I}) are reasonable approximations only if
$\Om_G\lsim\Om_B$. In the opposite case, which is
realized whenever
\bea
  \Om_B(\eta_{in}) &\gsim& \Om_{BG}(n)  \nonumber \\
 &\equiv& \left\{ \begin{array}{cc}
  {\Om_{\rm rad}\over
  12(n+3)}  & \mbox{for } n< -3/2 \\
  {\Om_{\rm rad}\over
  8(n+3)^2\log^2(\eta_{visc}/\eta_{in})}  & \mbox{for } n> -3/2 
 \end{array}\right. \nonumber \\
&=& \left\{ \begin{array}{cc} {3.3\times 10^{-6}h_0^{-2}\over
       (n+3)}  & \mbox{for } n< -3/2 \\
{5\times 10^{-6}h_0^{-2}\over
       (n+3)^2\log^2(\eta_{visc}/\eta_{in})}  & \mbox{for } n> -3/2~,
 \end{array}\right.   \label{OmBG}
\eea
the magnetic field energy is fully converted into gravity waves. 
Note, however, that the value $\Om_{BG}(n)$ is in general not very much 
smaller than $\Om_{\rm rad}$, which is an intrinsic limit on $\Om_B$ for our 
perturbative approach.

\begin{figure}[ht]
\centerline{\epsfxsize=3in  \epsfbox{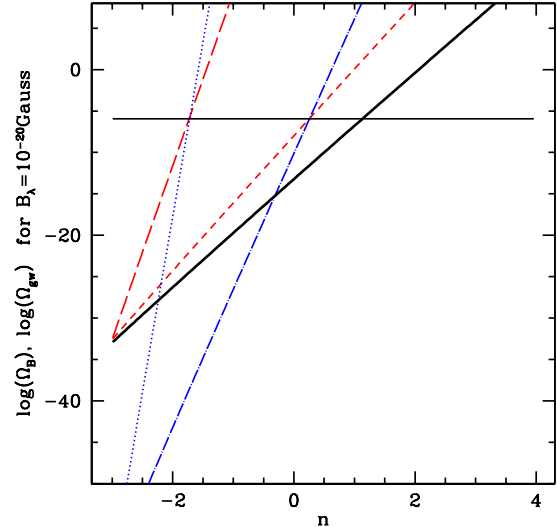}}
\caption{We show $\Om_Gh_0^2$ and $\Om_B(\eta_{in})h_0^2$ as functions of
the spectral index $n$ for two different  times of primordial magnetic
field creation: the electroweak transition ( $\Om_Gh_0^2$
dash-dotted, blue and $\Om_B(\eta_{in})h_0^2$ short-dashed, red), and
inflation  ( $\Om_Gh^2$ dotted, blue and $\Om_B(\eta_{in})h_0^2$
long-dashed, red) for a fiducial field strength $B_\la=10^{-20}$Gauss
at $\la=0.1$Mpc. The nucleosynthesis limit, $\Om_{\lim} h_0^2$ is also 
indicated. (The $\log$- terms have been neglected.) Clearly, the regimes 
with $\Om_B>1$ or $\Om_G>1$ are not physical and are just shown for 
illustration. We have also shown $\Om_B(\eta_{nuc})h^2$, the 
magnetic field density which is simply cut off at the 
nucleosynthesis damping scale (fat solid line).}
\label{OmBfig}
\end{figure}

In Fig.~\ref{OmBfig} the values
$\Om_G$ and $\Om_B(\eta_{in})$ as functions of the spectral index
are shown for two different choices of the creation time for the
primordial magnetic field: the electroweak transition,
$\eta_{in}=\eta_{ew}\sim 4\times 10^4$sec  and inflation with
$\eta_{in}\sim 8\times 10^{-9}$sec, for a magnetic field amplitude
$B_{\la} =10^{-20}$ Gauss. They are compared with the nucleosynthesis limit, 
which comes from the fact that an additional energy density may not change 
the expansion law during nucleosynthesis in a way which would spoil the 
agreement of the calculated Helium abundance with the observed value. 
The maximum allowed additional energy density is given by~\cite{michi}
\be \Om_{\lim}h_0^2 = 1.12\times 10^{-6} ~.
\ee

From Fig.~\ref{OmBfig} we see that $\Om_G$ as calculated above
dominates over $\Om_B(\eta_{in})$ for all spectral indices
 $n>-2$ in the inflationary case and $n>0$ for electroweak
magnetic field production, for an amplitude of $B_\la =10^{-20}$Gauss.
This is due to the fact that we
have neglected back-reaction which leads to a loss of magnetic
field energy. Clearly, the magnetic field cannot convert more than all
its energy into gravity waves. However, if  our formula for $\Om_G$ 
leads to $\Om_G>\Om_B(\eta_{in})$, it does actually
convert most of its energy into gravity waves, before it is dissipated by
plasma viscosity, since gravity wave production happens before and  at 
horizon crossing, while viscosity damping is active only on
scales which are well inside the horizon.  We can take into
account back-reaction by simply setting  $\Om_G \sim \Om_B(\eta_{in})$
when our calculation gives  $\Om_G>\Om_B(\eta_{in})$. We shall use 
this approximation for $\Om_G$ in what follows.

Fig.~\ref{OmBfig} also shows that, since the value of the magnetic
field density parameter at which conversion into gravity waves is
quasi complete is so close to the nucleosynthesis limit,
$\Om_{BG}(n)h_0^2 \sim 1.12\times 10^{-6} \equiv \Om_{\lim}h_0^2$, the two
curves $\Om_Gh_0^2$ and $\Om_B(\eta_{in})h_0^2$ cross close to
$\Om_{\lim}h_0^2$. This means that the gravity wave limit for magnetic
fields is very close to the limit obtained by setting  $\Om_G
=\Om_B(\eta_{in})$.

Let us discuss the problem of back-reaction in more detail.
Even if $\Om_G<\Om_B(\eta_{in})$, as soon as 
${d\Om_G(k)\over d\log (k)}> {d\Om_B(k)\over d\log (k)}$
for a given scale $k^{-1}$, we can no longer neglect back-reaction for 
this scale. The spectrum of  $\Om_G$ is  
\[ {d\Om_G(k)\over d\log (k)} \propto \left\{  \begin{array}{cc}
   k^{2n+6}, & \mbox{ for } n\le -3/2 \\  
   k^{3}, & \mbox{ for } n\ge -3/2, \\  
\end{array} \right.
\]
while  ${d\Om_B(k)\over d\log (k)} \propto k^{n+3}$.
Hence for $-3<n<0$, the gravity wave spectrum is 
bluer than the magnetic field spectrum. Since there is no infrared cutoff,
at sufficiently low values of $k$ we will always have 
${d\Om_G(k)\over d\log (k)}< {d\Om_B(k)\over d\log (k)}$ and back reaction 
is unimportant at low $k$. The value $k_{\lim}$, below which this is the case,
can be determined from  Eqs.~(\ref{OmBk},\ref{nl32k}) and (\ref{ng32k}). 
We find
\bea
\lefteqn{ k_{\lim}\la \left(\log^2(k_{\lim}\eta_{in})\right)^{1\over n+3}
  \simeq  \left({\Om_{\rm rad}\over
                   24\Om_\la}\right)^{1\over n+3}\sqrt{2}}
\nonumber \\
             &\sim&  [10^{26}(10^{-20}{\rm
Gauss}/B_\la)^2]^{1\over n+3}\sqrt{2} , \\
&& \mbox{ for } -3<n<-3/2 \nonumber \\
\lefteqn{ k_{\lim}\eta_{in} \simeq {1\over 2} \left({\sqrt{8}\Om_{\rm rad}\over
     24\Om_{in}\log^2(\eta_{visc}/\eta_{in})}\right)^{-1\over
        n}} \nonumber \\
&\sim&  \left[{2\times 10^{4}(10^{-9}{\rm Gauss}/B_{in})^2
   \over\log^2(\eta_{visc}/\eta_{in})}\right]^{-1\over n}~, \label{klim2}
 \\
& & \mbox{ for } -3/2<n<0, \nonumber 
\eea
where
\[  \Om_\la =B^2_\la/(8\pi\rho_c) \simeq 
	\left({d\Om_B(k)\over d\log(k)}\right)_{k=1/\la} ~~\mbox{ and}
\]
\[  \Om_{in}=\Om_\la(\eta_{in}/\la)^{n+3} \simeq	
	\left({d\Om_B(k)\over d\log(k)}\right)_{k=1/\eta_{in}}~,
\]
$B_{in}^2=B_\la^2(\la/\eta_{in})^{n+3}$.

If $k_{\lim}>1/\eta_{in}$, {\em e.g.} if the square bracket in 
Eq.~(\ref{klim2}) is larger than unity, back-reaction is never important.

For $n=0$ the magnetic field and gravity wave energy densities have the same
spectral index and the condition that gravity wave back-reaction becomes 
important is scale independent. In this case it simply reads
\be
 \Om_{in} \ge {\Om_{\rm rad} \sqrt{8} \over 24\log^2(\eta_{visc}/\eta_{in})}~.
 \label{limn0}	\ee 

The situation is different for $n>0$. Then the gravity wave spectrum is 
less blue than the magnetic field spectrum and back reaction is always 
important at sufficiently low $k$, large scales.

When back reaction is important, it leads to damping of the
primordial magnetic fields on large scales and will actually
damp the field down to values for which back-reaction is unimportant. 
This can be seen as follows: gravity wave production
takes place until $\Pi_{ij}(k)$, the tensor component of the magnetic field
stress tensor, vanishes. But then $f^2(k)=0$ which implies according to
Eq.~(\ref{ff})
\[ B^2(q) B^2(|{\bf k} -
        {\bf q}|)   =0~~~ \mbox{ for all}  ~~ 0\le q\le k_c .\]
For $n<0$
the quadratic nature of the coupling of $B$ to gravity waves actually
damps the magnetic field energy at least on all wave numbers
$q>k_{\lim}/2$.

For $n>0$, back-reaction reduces $\Pi_{ij}(k) \propto 
\int d^3q B^2(q) B^2(|{\bf k} -  {\bf q}|)$ for small enough values of 
$k$. In the limit $k\ra 0$, this indicates that back-reaction damps the 
magnetic field on {\bf all scales} until it becomes unimportant. 
It is difficult to decide without a detailed 
calculation how the magnetic field spectrum will actually be affected, but 
it seems reasonable to assume that back-reaction will alter it until 
$n\simeq 0$ and the amplitude until inequality (\ref{limn0}) is violated. 
We can therefore assume that in late time magnetic fields  inequality 
(\ref{limn0}) is always violated if the magnetic field spectral index is
$n\gsim 0$.

%We do not have strong observational limits on 
%$\Om_{in}$ from any other constraint than the one discussed here
%(except $\Om_{in}<1$). But clearly, for our perturbative calculation to 
%apply, we must require $\Om_{in}<\Om_{\rm rad}$. 
%A substantial 
%energy density in magnetic fields can not be excluded at very early time, 
%$\eta_{in}\ll \eta_{nuc}$. It may have disappeared into gravity waves or 
%by viscosity damping by the time of nucleosynthesis where we have reasonably
%stringent limits on unknown contributions to $\Om$. In contrary, the 
%magnetic field density from large scales, which has not significantly been 
%affected by gravity wave production or by viscosity damping until the time 
%of nucleosynthesis is constrained by the nucleosynthesis bound.

We find this a very important result, which can be summarized as follows:
Magnetic fields on super-horizon scale with a density which is sufficiently 
close to the radiation density are strongly damped into gravity waves when 
they enter the horizon. Note also that 'sufficiently close' can even mean 
several orders of magnitude smaller since $\log^2(k\eta_{in})$ can easily 
become of order 100 or more.
Furthermore, primordial magnetic fields produced on super horizon scales
have their spectral index changed by gravity wave production to $n\lsim 0$ 
once they enter the horizon.

During the matter dominated era gravity wave production is somewhat less 
efficient~\cite{RPT}; and since the scales of interest for us are sub-horizon
in the matter era we do not discuss it here.

\section{Limits and conclusions}

The first limit for primordial magnetic fields produced
before nucleosynthesis is simply
that the energy density which they contribute may not change
the expansion law during nucleosynthesis. As already mentioned,
this condition implies~\cite{michi}
\[ \Om_B(\eta_{nuc}) h_0^2 \le 1.12\times 10^{-6}
        = \Om_{\lim}h_0^2~.\]
Here we have disregarded the loss of magnetic field energy into gravity 
waves which will, as we shall see, strengthen the limit considerably.
From Eq.~(\ref{OmB}) we have
\bea
&& \Om_B(\eta_{nuc}) = {B_\la^2\over 8\pi\rho_c}
{(k_c(\eta_{nuc})\la)^{n+3}\over 2^{n+5\over 2}\Ga({n+5\over 2})}
\nonumber \\
&& ~~~ \simeq  {4.5h_0^{-2}\!\times\! 10^{-13}(5.9\!\times\!10^6)^n
\over 2^{n+5\over 2}\Ga({n+5\over 2})} \! \left({B_\la\over
10^{-20}{\rm G}}\right)^2\! \left({\la\over10^{13}{\rm sec}}\right)^{n+3}
\nonumber
\eea
where we have inserted\\  $k_d(\eta_{nuc})$  $\simeq$
$\sqrt{2\si_T\Om_b\rho_c/(\eta_{nuc}^3 m_{p}\Om_{\rm rad}H_0^2)}$ $\simeq$
$10^5/\eta_{nuc}$ $\simeq$  $6\times 10^{-7}$sec$^{-1}$ (for details
see Appendix~\ref{Apdamp} and Refs.~\cite{sblong,RPT,chiara}). The
density parameter $\Om_B(\eta_{nuc})$ as a function of the spectral index $n$ 
is shown in Fig.~\ref{OmBfig}.

\begin{figure}[ht]
\centerline{\epsfxsize=3in  \epsfbox{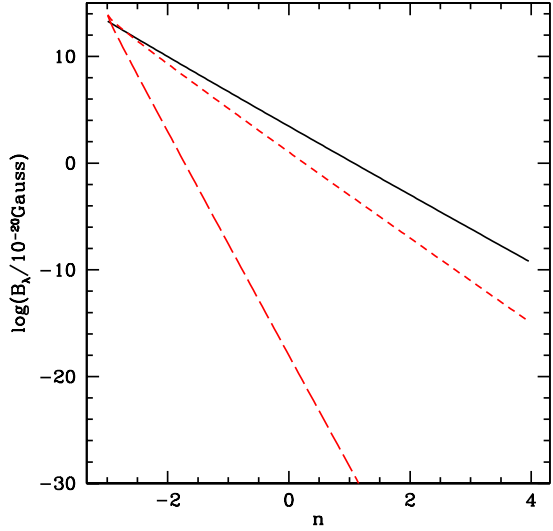}}
\caption{We show the nucleosynthesis limit on $B_\lambda$ (solid line)
as function of the spectral index, $n$ together with the limit from
gravity waves if the primordial field is produced at the electroweak
transition (short-dashed) or during inflation (long-dashed) for
$\lambda=0.1$h$^{-1}$Mpc$ \simeq 10^{13}$sec.}\label{Blim}
\end{figure}

Together with the above constraint, this gives already an interesting
limit on primordial magnetic fields with spectral indices $n>-2$, as
shown in Fig.~\ref{Blim} (solid line). For causal mechanisms of seed
field production, $n\ge 2$, it even implies $B_\la <10^{-22}$Gauss.

Nevertheless, the limit implied from the production of gravity waves
is more stringent, since the gravity waves have been produced at very
early times, when the magnetic field damping scale was much
smaller than $1/k_d(\eta_{nuc}) \sim 1.7\times10^6$sec. The production
of gravity waves has prevented the magnetic field energy from 
being lost by viscosity damping, since gravity waves do not interact 
with matter in any substantial way.

Setting $\Om_G = \Om_B(\eta_{in})$ whenever the result of
Eqs.~(\ref{nl32I},\ref{ng32I}) is larger than this limit, which is the
simplest way to account for back-reaction, the condition
\be \Om_Gh_0^2 < 1.12\times 10^{-6} = \Om_{\lim}h_0^2 \ee
yields the constraint for primordial magnetic fields created at
$\eta_{in}$.
For  spectral indices
\[ n> -3 +\sqrt{\Om_{\rm rad}\over 8\Om_{\lim}} \sim - 1 ~,\]
the value for  $\Om_G$ inferred from Eq.~(\ref{ng32I}) 
becomes larger than $\Om_B(\eta_{in})$ at
the limiting value $\Om_{\lim}$ imposed from nucleosynthesis
(in this approximation we have neglected the factor 
$\log^2(\eta_{visc}/\eta_{in})$, which can be considerable!). Then the 
magnetic field damping due to gravity wave productions is very important.
But also for smaller values of the spectral index, $n>-3$, we have $\Om_G 
\sim
\Om_B(\eta_{in})$  for $\Om_G \sim \Om_{\lim}$ and there is still
a considerable amount of magnetic field damping due to gravity wave 
production.

The results for primordial magnetic fields produced at
inflation and at the electroweak scale are shown in Fig.~\ref{Blim}
(dashed lines). As can be seen for the two examples,
primordial magnetic fields produced before nucleosynthesis are very 
strongly constrained. For all values of the spectral index, the following 
expression is a good  approximation for the limit obtained:
\bea
   B_\la/10^{-9}{\rm Gauss} &<&700h_0\times(\eta_{in}/\la)^{(n+3)/2}{\cal
        N}(n)   \label{Bbound}\\
\mbox{where }~ {\cal N}(n)  &\equiv&
   \sqrt{2^{n+5\over 2}\Ga\left({n+5\over 2}\right)} \sim 1 ~. \nonumber
\eea
This nucleosynthesis bound becomes
stronger for smaller cutoff scales, larger $k_c$, according to
Eq.~(\ref{Bbound}) it scales like $(k_c\la)^{-(n+3)/2}$. (Remember that we 
have set $k_c=1/\eta_{in}$.)

If the seed field is produced
during an inflationary phase at GUT scale temperatures, where conformal
invariance can be broken {\it e.g.} by the presence of a dilaton, the
induced fields must be smaller than $B_\la \sim 10^{-20}$Gauss for
$n>-2$. If seed fields are produced after inflation, their spectrum is
constrained by causality. Deviation from a power law with
$n\ge 2$ can only be  produced on sub-horizon scales,
$k>1/\eta_{in}$. Therefore our limit derived by setting $B(k)=0$ on
sub-horizon scales, $k\eta_{in}>1$, is the most
conservative choice consistent with causality.

Mechanisms which still can produce significant seed fields are
either 'ordinary' inflation, if the spectral index $n\lsim -2$ or a late
inflationary phase at the electroweak scale (or even later) where a seed 
field
with $n\lsim 0$ can have amplitudes of $B_\la \sim 10^{-20}$Gauss.

We also have found that magnetic fields which contribute
an energy density  close to the nucleosynthesis bound, loose a
considerable amount (if not all) of their energy into gravity waves,
which might be detectable. In fact, the space born interferometer
approved by the European Space Agency and NASA, the Large Interferometer 
Space Antenna (LISA) which has its most
sensitive regime where it can detect $\Om_Gh_0^2 \sim 10^{-11}$
around $10^{-3}$Hz$\sim 1/\eta_{weak}$~\cite{michi} will either
detect or rule out all magnetic seed fields with spectral index $n\gsim
-0.5$ produced around
or before the electroweak phase transition. If LISA does not detect 
a
gravity wave background, the  constraint analogous to
Eq.~(\ref{Bbound}) for $\eta_{in} \le 4\times 10^4$sec yields
\[ B_\la < 10^{-20}{\rm Gauss} ~~\mbox{ for all indices } n> -0.5 \]
for all mechanisms producing seed fields before or at the electroweak
phase transition.

We  conclude that, most probably, magnetic seed fields have to be 
produced relatively
late, or after nucleosynthesis to evade the discussed bounds.
Our gravity wave bound is not relevant for magnetic fields which are
produced on sub-horizon scales. But for  $\la\gsim
0.1$Mpc to enter the horizon, this requires a temperature of creation  $T 
< 1$keV. The only late time mechanism found so far
which could lead to seed fields is recombination, where large scale
 fields of the order of $B\sim 10^{-20}$ Gauss can be
induced by magneto-hydrodynamic effects, and the difference in the
viscosity of electrons and ions~\cite{craig}, a charge separation
mechanism. Our work strongly constrains processes of quantum
particle production (during {\it e.g.} an inflationary phase) as
origin for the observed magnetic fields and favors more
conventional processes like charge separation in the late universe.

\vspace{0.3cm}

\noindent {\bf Acknowledgment:} \hspace{2mm} We thank Pedro Ferreira,
Michele Maggiore and Roy Maartens for helpful
discussions. This work is supported by the Swiss NSF.

\vspace{1.3cm}

\appendix

\section{Damping of magnetic fields by viscosity}
\label{Apdamp}
In this appendix we  determine the cutoff function $k_{d}(\eta)$.
We use the results found in~\cite{sblong,mhd} and \cite{chiara}.

We split the magnetic field into a high frequency and a low
frequency component, separated by the Alfv\'en
scale, $\lambda_{A}=v_{A}\eta$, where the Alfv\'en velocity
\[  v_A^2 = {\langle B^2\rangle\over 4\pi(\rho_r+p_r)}\]
 depends
on the low frequency component: $\langle B^2_{A}\rangle=\langle B_{0i}({\bf
x})B_{0}^{i}({\bf x})\rangle|_{\lambda_{A}}$, 
$v_A \sim 4\times 10^{-4}\times (B_A/10^{-9}{\rm Gauss})$ 
~\cite{RPT}.
The amplitude of the high frequency component then obeys a damped
harmonic oscillator equation, with damping coefficient, $D(\eta)$,
depending on time and on the
mean free path of the diffusing particles giving rise to 
viscosity~\cite{sblong}. In the oscillatory regime, we define the damping scale
at each time $\eta$
to be the scale at which one e-fold of damping has occurred:
$\int_{0}^{\eta}\frac{D}{2}d\eta=1$.
The damping term $D$ is given by $D=k^2\la_{col}/a(\eta)$, where $\la_{col}$ 
is the
mean free path of the particle species with the highest viscosity which is
still sufficiently strongly coupled to the magnetic field.
Long wave modes with $1/k>v_A\eta$ are not 
significantly damped. We now determine the damping scale as a function of 
time. To determine whether a given mode with $k>k_d(\eta)$ is effectively 
damped one has to decide whether it is in the oscillatory regime, 
$\om_0=kv_A > D = k^2\la_{col}/a(\eta)$ where damping really has time to 
occur or in the 'over-damped' regime  $k< v_Aa(\eta)/2\la_{col}$ where 
amplitudes remain approximatively constant. With $v_A$ this depends on the 
magnetic field under consideration.

Let us now determine the damping scale.
Before neutrino decoupling at $T\gsim 1$MeV corresponding to
$\eta\lsim 10^{10}$sec, damping is due to both photon and
neutrino viscosity. The mean free path of photons is
$$\la_{col,\gamma}\simeq \frac{1}{\sigma_{T}n_{e}}\simeq a^3 (1.5\times10^{20}
\mathrm{sec})~,$$
where $\sigma_{T}=6.65\times 10^{-25}{\rm cm}^2$ is the cross section of 
Thomson scattering.
For neutrinos, we take into account scattering with leptons as
the principle scattering  process giving rise to viscosity:
$$\la_{col,\nu}\simeq\frac{1}{\sigma_{w}n_{\nu}} \simeq a^5 
	(7\times10^{48}\mathrm{sec}) ~,$$
where $\si_w =G_F^2T^2$ is the weak cross section and 
$G_F= (293{\rm GeV})^{-2}$ is Fermi's constant. Note that we set $\hbar=c=1$
so that a cross section also can have the units GeV$^{-2}$.

Using the expression for the scale factor given in Eq.~(\ref{scfac}), 
one finds that photon viscosity dominates until 
$\eta\simeq10^5$ sec, leading to
\be
k_{d}(\eta)\simeq (2\times10^{10}\mathrm{sec}^{1/2})\eta^{-3/2}~.
\label{dampph}
\ee 
For $\eta>10^5$ sec neutrinos viscosity takes over, with
cutoff function
\be
k_{d}(\eta)=(4\times10^{15}\mathrm{sec}^{3/2})\eta^{-5/2}
\label{dampnu}
\ee 
during the oscillatory regime.
The comoving wavenumber $k$ is given here in units of 
sec$^{-1}$. 

After $\eta\gsim 10^{10}$sec neutrinos decouple and the dominant viscosity 
is again photon viscosity leading to the cutoff function (\ref{dampph}).

Estimating the viscosity time, namely $k_d(\eta_{visc})=1/\eta_{in}$ for
inflation, $\eta_{in}\sim 10^{-8}$sec and the electroweak phase transition,
$\eta_{in}=\eta_{ew}\simeq 4\times 10^4$sec, we find from the expressions 
above 
$ \eta_{visc}/\eta_{in}|_{\rm inflation} \sim 3\times 10^9 $
and
$ \eta_{visc}/\eta_{ew} \sim 3000. $
The first result is calculated using photon viscosity is just 
approximative, since
we do not know the relevant cross sections up to the scale of inflation, 
$10^{15}$GeV, but we certainly expect the value to be very large, since 
interactions are strong and thus viscosity is weak. The electroweak result,
calculated using the neutrino viscosity, would be quite reliable in the 
oscillatory regime. However, for magnetic fields $B<10^{-9}$Gauss, for which
the Alfv\'en velocity is smaller than $10^{-4}$, the scale
$\eta_{visc}$ is still in the over-damped regime. The time
at which the scale can then effectively be damped depends on the value of 
the magnetic field. In this sense our result is only a lower limit,
$ \eta_{visc}/\eta_{ew} \gsim 3000. $ This is not very important for our
final bounds, where we will even set $\log{ \eta_{visc}/\eta_{in}} \sim 1$,
in order to obtain results which are independent of the time of magnetic field 
creation.  

As an example we also determine the damping scale at nucleosynthesis, 
$T\simeq 0.1$MeV, $z_{nuc}\simeq 4\times 10^8$ which we  need in Section~4.
Setting $D\eta/2=1$, we obtain
\be 
 k_d(\eta_{nuc}) = [2a(\eta_{nuc})\si_Tn_e(\eta_{nuc})/\eta_{nuc}]^{1/2}~. 
\ee
Using $n_e =\rho_c\Om_b/(m_pa^3)$, where $m_p$ is the proton mass, as well 
as our  expression for the scale factor one obtains
\[ k_d(\eta_{nuc}) \simeq 6\times 10^{-7}{\rm sec}^{-1}\simeq 10^5/\eta_{nuc}
\]
This can of course also be obtained by simply using 
$\eta_{nuc}\simeq 10^{11}$sec in the above function for photon 
viscosity given in Eq.~(\ref{dampph}). Again, whether or not this scale is 
in the oscillatory regime and can be effectively damped, depends on the value
of $B(k_d)$. For $B(k_d)\sim 10^{-6}$Gauss, which satisfies the nucleosynthesis
bound, this is largely the case, and for magnetic fields of interest to us
$ k_d(\eta_{nuc})$ is the correct damping scale.

At the end of the radiation dominated era, photons decouple and viscosity acts
no more. Since gravity wave production in the matter dominated regime is 
not important, we do not calculate the cutoff function in this regime.

\section{The gravity wave source of stochastic magnetic fields}
\label{Apten}
The Maxwell stress tensor of a  magnetic field in real 
space  is given by
\bean T^{ij}(\bx,\eta)&=&{1\over 4\pi}\big[ B^i(\bx,\eta)B^j(\bx,\eta) \\
	&& -{1\over 2}
	g^{ij}(\bx,\eta)B_n(\bx,\eta)B^n(\bx,\eta)\big]~.
\eean
In Fourier space, using the Fourier transform convention adopted in this paper
and the scaling of the magnetic field with time, we have
 \bea
T^{ij}({\bf k},\eta) &=& {1\over 4\pi(2\pi)^3 a^6}\int d^3q
\big[B^{i}({\bf  q})B^{j}({\bf
        k-q}) \nonumber \\
&&  - {1\over 2}B^{l}({\bf q})B^{l}({\bf k-q})\de^{ij}\big] ~,
\label{tauapp}
\eea 
where  we have introduced the factor $1/a^6$ to transform the
present field $B^i(\bk)= B^i(\bk,\eta_0)$ back to the physical field
$B^i(\bk,\eta)=B^i(\bk)/a^3$. 
$\Pi^{ij}(\bk,\eta)$ is the transverse traceless component of 
$T^{ij}(\bk,\eta)$, which sources gravity waves.
Here we give the details of the calculation of its correlation function,
$\langle\Pi^{ij}(\bk,\eta)\Pi^{*lm}(\bk'\eta)\rangle$ which we use to 
compute the induced gravity waves.
The projector onto the component of a vector transverse to $\bk$ is
$P_{ij} = \de_{ij}-\hat{k}_i\hat{k}_j$. Consequently
$P^{i}_{a}P^{j}_{b}$ projects onto the transverse 
component of a tensor. To obtain the transverse traceless component we
still have to subtract the trace. Hence defining the projector
\[{\mathcal P}^{ij}_{ab}=P^{i}_{a}P^{j}_{b}-{1\over 2}P^{ij}P_{ab}   
\]
we have
\begin{equation}
\langle\Pi^{ij}(\bk,\eta)\Pi^{*lm}(\bk 
',\eta)\rangle={\mathcal P}^{ij}_{~~ab}{\mathcal P}^{lm}_{~~cd}
\langle T^{ab}(\bk,\eta)T^{*cd}(\bk ',\eta)\rangle\: .
\label{P1ij}
\end{equation}
To simplify the calculation, we note that up to a trace, which anyway 
vanishes in the projection (\ref{P1ij}), $T^{ab}(\bk,\eta)$ is just given by
\be \De^{ab}(\bk,\eta) \equiv  {1\over 4\pi(2\pi)^3 a^6}\int d^3q
  B^{a}({\bf  q})B^{b}({\bf k-q})~. \label{ApDe}
\ee
We therefore can write
\begin{equation}
\langle\Pi^{ij}(\bk,\eta)\Pi^{*lm}(\bk 
',\eta)\rangle={\mathcal P}^{ij}_{~~ab}{\mathcal P}^{lm}_{~~cd}
\langle\De^{ab}(\bk,\eta)\De^{*cd}(\bk ',\eta)\rangle\: .
\label{Pij}
\end{equation}
To compute the two point correlator of $\De$, we use expression 
(\ref{ApDe}) and the assumption that the random magnetic field be Gaussian, 
so that we can apply Wick's theorem. In other words, products of four magnetic 
fields can be reduced by
\bea
\lefteqn{\langle B^i({\bf k})B^{*j}({\bf q}) B^n({\bf s})B^{*m}({\bf p}
)\rangle   = ~~~~~~~~~~} \nonumber \\  &&
~~~~~~~~~~\langle B^i({\bf k})B^{*j}({\bf q})\rangle
\langle B^n({\bf s})B^{*m}({\bf p})\rangle + \nonumber \\ &&
~~~~~~~~~~ \langle B^i({\bf k})B^{n}({\bf s})
\rangle\langle B^{*j}({\bf q})B^{*m}({\bf p})\rangle + \nonumber \\ &&
~~~~~~~~~~ \langle B^i({\bf k})B^{*m}({\bf p})\rangle
\langle B^n({\bf s})B^{*j}({\bf q})\rangle ~. \label{corB2}
\eea

 Using also the reality condition, $B^{*a}(\bk)=B^a(-\bk)$, and the 
two point correlator (\ref{powB}), we obtain
\begin{eqnarray}
& &\langle\De^{ab}(\bk,\eta)\De^{*cd}(\bk 
',\eta)\rangle=\frac{a^{-12}}{4(2\pi)^8}\int d^3q d^3p
[\delta(\bk)\delta(\bk ')\times \nonumber \\
 && B^2(q)B^2(-p)(\delta^{ab}-\hat{q}^{a}\hat{q}^b)
(\delta^{cd}-\hat{p}^{c}\hat{p}^d)+ \nonumber \\
& &+\delta(\bq-\bp)\delta(\bk-\bq-\bk 
'+\bp)B^2(q)B^2(|\bk-\bq|)\times \nonumber \\
&& ~~~(\delta^{ac}-\hat{q}^{a}\hat{q}^c)
(\delta^{bd}-(\widehat{\bk-\bq})^{b}(\widehat{\bk-\bq})^d)+ \nonumber \\
& &+\delta(\bq-\bk'+\bp)\delta(\bk-\bq-\bp)B^2(q)B^2(|\bk-\bq|) 
  \times \nonumber \\
&& ~~ (\delta^{ad}-\hat{q}^{a}\hat{q}^d)
(\delta^{bc}-(\widehat{\bk-\bq})^{b}(\widehat{\bk-\bq})^c)].
\label{tauab}
\end{eqnarray}
The first term only contributes an uninteresting constant and can 
be disregarded. For the remaining two terms integration over $d^3p$ 
eliminates one of the two $\delta$-functions and leads to
\bea
\lefteqn{\langle\De^{ab}({\bf k},\eta)\De^{*cd}({\bf
k'},\eta)\rangle =} \nonumber\\
 && \delta({\bf k}-{\bf k'}){a^{-12}\over 4(2\pi)^8} 
\int d^3 q~B^2(q) B^2(|{\bf k}-{\bf q}|)
\times  \nonumber \\
 && \left[ (\delta^{ac}- {\hat q}^a {\hat q}^c)
(\delta^{bd}- {\widehat {({\bf k}-{\bf q})}}^b {\widehat
{({\bf k}- {\bf q})}}^d)+  \right.
     \nonumber   \\
 && \left. (\delta^{ad}- {\hat q}^a{\hat q}^d)
(\delta^{bc}- {\widehat{({\bf k}-{\bf q})}}^b
        { \widehat{({\bf k}-{\bf q})}}^c) \right]  ~.
	  \label{ijlmapp}
\eea
Clearly, the correlator of $\De$ and thus also the one of $\Pi$ is 
symmetric in $\bk$ and $\bk'$ and hence also under the exchange of the first
and the second pair of indices. In addition it is symmetric in the first 
and the second as well as in the third and the fourth index. The most 
general isotropic transverse traceless fourth rank tensor which obeys 
these symmetries has the tensorial structure
\bea
{\cal M}^{ijlm}(\bk)&=& \de^{il}\de^{jm}+\de^{im}\de^{jl}  
-\de^{ij}\de^{lm} +
        k^{-2}(\de^{ij}k^lk^m + \nonumber \\
&&      \de^{lm}k^ik^j -\de^{im}k^jk^l - \de^{il}k^jk^m -\de^{jl}k^ik^m
        \nonumber \\
&&      -\de^{jm}k^ik^l) + k^{-4}k^ik^jk^lk^m~.
\eea
We could not find a straight forward derivation of this result in a textbook
on multi-linear algebra where it actually belongs, but it can be found, 
e.g. in~\cite{DK}.

We can hence set 
\bea
\langle \Pi^{ij}({\bf k},\eta)\Pi^{*lm}({\bf
        k}',\eta) \rangle &=&f(k,\eta)^2/a^{12}{\cal M}^{ijlm}\de({\bf
k-k'})\nonumber \\
\mbox{ with} && \nonumber  \\
\langle \Pi_{ij}({\bf k},\eta) \Pi^{*ij}({\bf k'},\eta)\rangle
&=& {4\over a^8}f(k,\eta)^2\de({\bf k-k'}),
\label{AppowPi}
\eea
To determine the correlator of $\Pi$ it is therefore sufficient to calculate 
its trace.
With  ${\mathcal P}_{ijab}{\mathcal P}^{ij}_{cd} =
   {\mathcal P}_{abij}{\mathcal P}_{cd}^{ij} ={\mathcal P}_{abcd}$, (for the 
last identity we simply use that projectors are idem-potent), we have
\be
\langle \Pi_{ij}({\bf k},\eta) \Pi^{*ij}({\bf k'},\eta)\rangle =
 {\mathcal P}^{abcd}\langle 
  \De_{ab}({\bf k},\eta) \De_{*cd}({\bf k'},\eta)\rangle~.
\ee
A somewhat tedious but straight forward computation gives
\begin{eqnarray}
& &{\mathcal P}^{abcd}[(\delta^{ac}-\hat{q}^{a}\hat{q}^c)
(\delta^{bd}-(\widehat{\bk-\bq})^{b}(\widehat{\bk-\bq})^d) \nonumber \\
& & ~~ \qquad +(\delta^{ad}-\hat{q}^{a}\hat{q}^d)
(\delta^{bc}-(\widehat{\bk-\bq})^{b}(\widehat{\bk-\bq})^c)]= \nonumber \\
& &1+(\hat{\bk}\!\cdot\!(\widehat{\bk\!-\!\bq}))^2+
(\hat{\bk}\!\cdot\!\hat{\bq})^2+
(\hat{\bk}\!\cdot\!\hat{\bq})^2
(\hat{\bk}\!\cdot\!(\widehat{\bk\!-\!\bq}))^2\:.
\label{alg}
\end{eqnarray}
Setting $\gamma=\hat{\bk}\cdot\hat{\bq}$ and
$\beta=\hat{\bk}\cdot(\widehat{\bk-\bq})$, and using the fact that the 
second term transforms into the third one under the transformation 
$\bq \rightarrow \bk-\bq$, we finally obtain
%$$\int d^3q B^2(q)B^2(|\bk-\bq|)(\hat{\bk}\cdot(\widehat{\bk-\bq}))^2=
%\int d^3u B^2(u)B^2(|\bk-\mathbf{u}|)(\hat{\bk}\cdot\hat{\mathbf{u}})^2$$
%if $\mathbf{u}=\bk-\bq$, we finally obtain
\bea
&& \langle\Pi_{ij}(\bk,\eta)\Pi^{*ij}(\bk 
',\eta)\rangle=\frac{a^{-8}}{4(2\pi)^8}
\delta(\bk-\bk)\times \nonumber \\  &&  \qquad \int d^3q 
B^2(q)B^2(|\bk-\bq|)(1+2\gamma^2+\gamma^2\beta^2)\:,
\label{Pijfin}
\eea
which leads to the result for $f(k)$ given in Eq.~(\ref{ff}).

\section{Gravitational wave production}
\label{apwave}
The equation for gravity wave production due to tensor type anisotropic
stresses is
\be
{\ddot{h}}_{ij}+2{\dot a \over a}\dot{h}_{ij}+
k^2 h_{ij}=8\pi G \Pi_{ij} ~.
\ee
For each mode we therefore have an equation of the form
\be
{\ddot{h}}+2{\dot a \over a}\dot{h}+ k^2 h= s(k,\eta)~, \label{Aphprime}
\ee
where $s(k,\eta)={8\pi G\over a^2}f(k,\eta)$. The function $f$ only depends
on $\eta$ for $n>-3/2$ via the damping cutoff $k_d(\eta)$.
In terms of the dimensionless variable $x=k\eta$ equation (\ref{Aphprime}) 
reduces to
\be
h''+2{\al \over x} h'+ h= s(k,\eta)k^{-2}~, \label{Aphprimex}
\ee
where $\al=1$ in the radiation dominated era, and $\al=2$ in the matter
dominated era. The homogeneous solutions of Eq.~(\ref{Aphprimex}) are the  
spherical Bessel functions $j_0~,y_0$ in the radiation  dominated era,
and $j_1/x,~y_1/x$ in the matter dominated era respectively.
We assume that the magnetic fields were created in the radiation
dominated epoch, at redshift $z_{in}$. Using the  Wronskian method, the
general solution of Eq.~(\ref{Aphprimex}) which vanishes at $z_{in}$ is
given by
\be
h(x) = c_1(x)g_1(x) +c_2(x)g_2(x)~,
\ee
where $g_1,g_2$ are the above mentioned homogeneous solutions and
\bean
c_1(x) &=& -k^{-2}\int_{x_{in}}^x s(x')g_2(x')/W(x')dx' \\
c_2(x) &=& k^{-2}\int_{x_{in}}^x s(x')g_1(x')/W(x')dx'~, \\
\eean
$W =g_1g_2'-g_1'g_2$ is the Wronskian determinant of the homogeneous
solution. Inside the horizon the homogeneous solutions $g_1$ and $g_2$ begin
to oscillate. The contribution to the integral from times where the scale
under consideration is sub-horizon is hence negligible. Furthermore, since the
gravity wave energy is growing with wave number (it is proportional to
$k^3f^2$), our limit will come from large wave numbers, small scales,
which enter the horizon before decoupling. Let us thus solve
Eq.~(\ref{Aphprimex}) explicitly in the radiation dominated regime,
$\eta<\eta_{eq}$, for a wave number which enters the horizon in the radiation
era, $k\eta_{eq}>1$, and in the case where $f$ is not time dependent 
($n<-3/2$). We first notice that the Wronskian $W(j_0,y_0)=1/x^2$.
Using the radiation approximation of~
Eq.~(\ref{scfac}) for the scale factor, $a=H_0\eta\sqrt{\Om_{\rm rad}}$
we have
\[  {k^{-2}s(x)\over W(x)} = {8\pi G f(k)\over H_0^2\Om_{\rm rad}}~.
\]
Since $y_0$ diverges at small $x$ the term $c_1$ clearly dominates.
After horizon crossing we have
\[  h(x) \simeq c_1(1)j_0(x) = c_1(1){\sin x\over x} ~.
\]
Performing the integral $c_1(1)$, we find
\be
h(x) \simeq - {8\pi Gf(k)\over H_0^2\Om_{\rm rad}}{\sin x \over x}
        \log(x_{in}) ~,
        \label{aphx}
\ee
for $x>1$ and $\eta<\eta_{eq}\simeq \sqrt{\Om_{\rm rad}}/H_0$.
We have compared this formula with the numerical solution and, as expected,
found that it is a very reasonable approximation (within less than 10\% of the
numerical result).

After horizon crossing, the gravity waves thus propagate freely, and their
energy just scales like radiation energy, so that for $k\eta_{eq}>1$, using
Eq.~(\ref{Omgk})
\bea
{d\Om_G(k)\over d\log(k)} &\simeq& {d\rho_G(k)\over\rho_{\rm rad}\,d\log 
(k)}
\Om_{\rm rad}
  = {k^3\dot h^2 \over a^2\rho_{\rm rad}(2\pi)^6G}\Om_{\rm rad}~.
\eea
During the radiation era, on sub-horizon scales
\[
\dot h \simeq {8\pi Gf(k)\over\eta H_0^2\Om_{\rm rad}}\log(x_{in})\cos(x)
\mbox{ and }
a^2\rho_{\rm rad} = {3\over 8\pi G}\left({1\over \eta}\right)^2
\]
so that
\bea
{d\Om_G(k)\over d\log(k)} &=& {4k^3f(k)^2(8\pi G)^2\log^2(x_{in})\cos^2(x)
   \over H_0^4\Om_{\rm rad}3(2\pi)^5} \nonumber \\
 &\simeq& {12k^3f(k)^2\log^2(x_{in})
\over \rho_c^2\Om_{\rm rad}(2\pi)^5}~.
\eea
Since the ratio between the gravity wave energy density and the
radiation energy density is time independent, this formula is valid also
in the matter era. $\rho_c=3H_0^2/(8\pi G)$ denotes the critical
density today.

\end{document}